# THE ART OF COMMUNITY DETECTION

NATALI GULBAHCE AND SUNE LEHMANN

Center for Complex Networks Research, Northeastern University, Boston, MA 02115, USA, and Center for Cancer Systems Biology, Dana Farber Cancer Institute, Boston, MA 02215, USA.

## SUMMARY

Networks in nature possess a remarkable amount of structure. Via a series of data-driven discoveries, the cutting edge of network science has recently progressed from positing that the random graphs of mathematical graph theory might accurately describe real networks to the current viewpoint that networks in nature are highly complex and structured entities. The identification of high order structures in networks unveils insights into their functional organization. Recently, Clauset, Moore, and Newman[1], introduced a new algorithm that identifies such heterogeneities in complex networks by utilizing the hierarchy that necessarily organizes the many levels of structure. Here, we anchor their algorithm in a general community detection framework and discuss the future of community detection.

## STRUCTURE EVERYWHERE

The view that networks are essentially random was challenged in 1999 when it was discovered that the distribution of number of links per node (degree) of many real networks (internet, metabolic network, sexual contacts, airports, etc) is different from what is expected in random networks[2]. In a large random network node degrees are distributed according to the normal distribution, but in many man-made and biological networks the degree distribution follows a power-law. In the human protein-protein interaction networks[3,4], for instance, some proteins act as hubs, they are highly connected, and interact with more than 200 other proteins contrary to most proteins that interact with only a few other proteins.

Various local to global measures have been introduced to unveil the organizational principles of complex networks[5,6,7,8,9]. Maslov and Sneppen[10] discovered that who links to whom can depend on node degree; in many biological networks, high degree nodes systematically link to nodes of low degree. This disassortativity decreases the likelihood of cross talk between functional modules inside the cell and increases overall robustness. Other networks, for example social networks[11], are highly assortative – in these networks nodes with similar degree tend to link to each other.

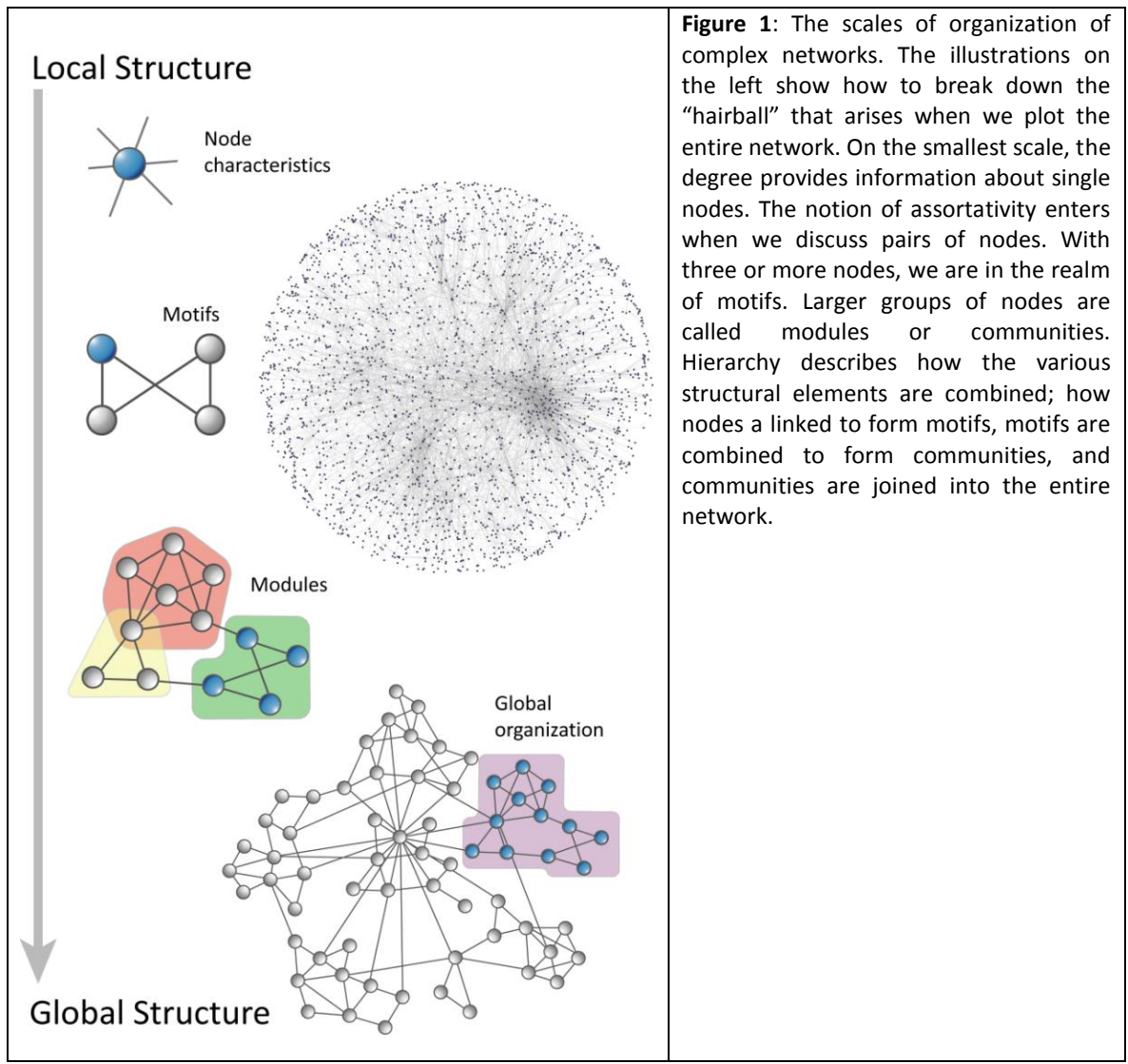

**Figure 1**: The scales of organization of complex networks. The illustrations on the left show how to break down the "hairball" that arises when we plot the entire network. On the smallest scale, the degree provides information about single nodes. The notion of assortativity enters when we discuss pairs of nodes. With three or more nodes, we are in the realm of motifs. Larger groups of nodes are called modules or communities. Hierarchy describes how the various structural elements are combined; how nodes a linked to form motifs, motifs are combined to form communities, and communities are joined into the entire network.

Going beyond the properties of single nodes and pairs of nodes, the natural next step is to consider structures that include several nodes. Interestingly, a few select motifs of three to four nodes are ubiquitous in real networks[12] while most others occur only as often as they would at random, or, are actively suppressed. Other local measures that signify a dense or sparse local structure in a network are the clustering coefficient[13] and short loops[14,15].

## COMMUNITIES

Between the scale of the whole network and the scale of the motifs we find the network communities[16,17]. A community is a densely connected subset of nodes that is only sparsely linked to the remaining network. Modular structure introduces important heterogeneities in complex networks. For example, each module can have different local statistics[18]; some modules may have many connections, while other modules may be sparse. When there is large variation among communities, global values of statistical measures can be misleading. The presence of modular structure may also alter the way in which dynamical processes (e.g., spreading processes and synchronization[19]) unfold on the network. In biological networks, communities correspond to functional modules in which members of a module function coherently to perform essential cellular tasks. Both metabolic networks[20] and protein phosphorylation networks[21], for example, possess high clustering coefficients and are modular.

The ultimate goal in biology is to determine how genes and the proteins they encode function in the cell. A revolutionary approach to discover gene function has been to knock out a gene and observe its phenotype. A nearly complete collection of single gene deletions has been performed for *Saccharomyces Cerevisiae*[22]. Eukaryotes show large amounts of genetic redundancy, however, and single knock outs are no longer informative. Hence, the function of a large number of genes remains unknown[23]. The deletion of multiple genes can yield a wealth of information about gene function and epistasis[24,25]. Despite the advancing experimental techniques[23], systematic multiple gene deletions for more than two genes quickly become impossible due to the high number of possible gene combinations.

A promising computational approach to discovery of functions of genes and proteins is to identify functional modules in biological networks. Since modules are sets of genes or proteins that perform biological processes together, it is possible to classify proteins with unknown function by determining what module they belong to[26]. Correct identification of functional modules also has important biotechnological and drug design applications. In many cases the deletion of a certain function may be necessary and this can be achieved by removing the entire functional module.

A standard method for detecting modules in complex networks has yet to be agreed upon[16]. One popular approach sees communities as sets of adjacent motifs[26], other methods are inspired by information theory[27], message passing[28], or Bayesian principles[29-30]. A widely used class of algorithms is based on optimization of a quantity called *modularity*[31]. The modularity is proportional to the difference between the number of edges within communities and the expected number of such edges. As we shall see in the following, the algorithm by Clauset et al. is more general than other community detection algorithms because it is able to infer other structures than communities from the network data.

HIERARCHY

Hierarchy describes the organization of elements in a network; how nodes link to each other to form motifs, how motifs combine to form communities and how communities are joined to form the entire network (see Figure 1). Data clustering algorithms that find successive modules using previously established modules are termed hierarchical clustering[32]. Hierarchical clustering has a long history in biology[33], e.g., to find coexpressed genes or to assign genotypes in high throughput genotyping platforms.

Ravasz et al.[20] have shown that the metabolic network of several organisms can be organized into highly connected modules that hierarchically combine into larger units. In particular, within *Escherichia coli*, the observed hierarchy coincides with known metabolic functions. Incorporating hierarchy into graph theoretical models allows the simultaneous description of two distinct systems-level features of real world networks, power-law degree distribution and modular topology[34].

## Hierarchical Random Graphs

The basis for the algorithm suggested by Clauset et al. is the *hierarchical random graph model*. In a (non-hierarchical) random graph model with *n* nodes, each pair of nodes is connected with the same probability *p*. On average, each realization of such a network has $p \cdot n(n-1)/2$ links. In the hierarchical random graph model, the probability of two nodes connecting is not a constant *p*, but rather a hierarchy of probabilities (see Figure 2 for details).

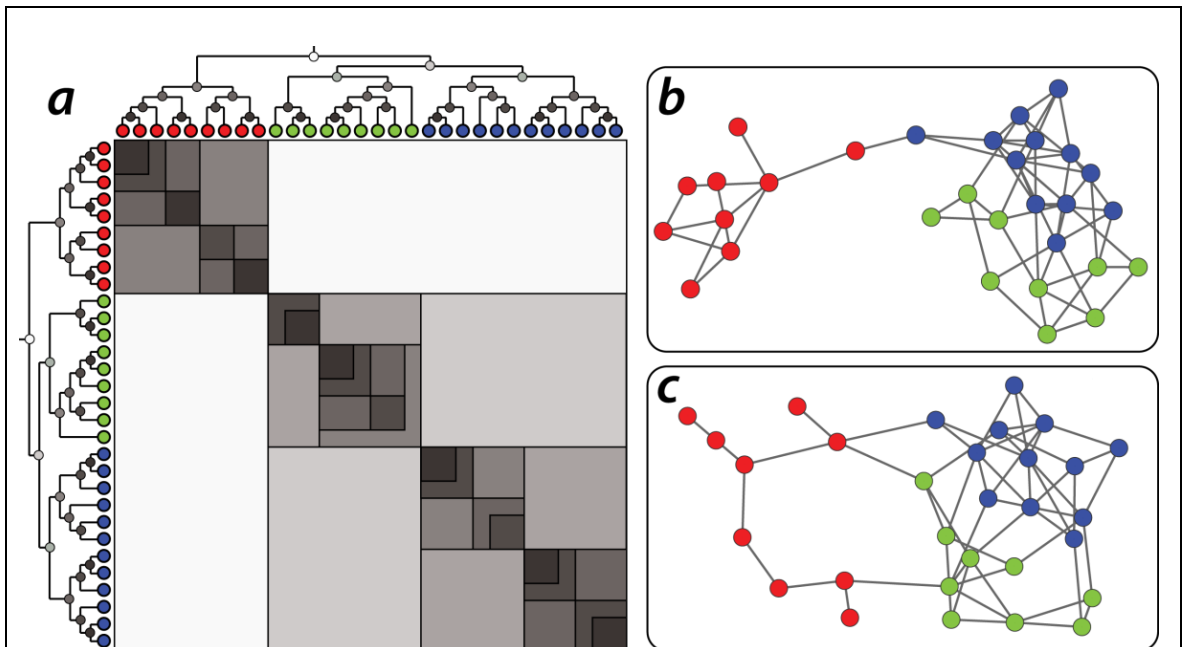

**Figure 2**. The hierarchical random graph model. Panel *a*. shows an example of a hierarchical random graph with a community-like structure. We illustrate the connection between the dendrogram

representation used by Clauset et al. and an equivalent adjacency matrix representation of the hierarchical random graph model. The leaves of the dendrogram are the 30 network nodes (red, green, and blue are used to visually suggest the three major divisions of the model). Each internal node $r$ encodes the probability $p_r$ of connecting each pair of vertices for which $r$ is the lowest common ancestor. The shading of internal nodes corresponds to probabilities on a scale where black is $p_r = 1$ and white is $p_r = 0$. The model is also visualized as a matrix $A$ of probabilities, where matrix element $A_{ij}$ describes the probability of a link between nodes $i$ and $j$. The shading matches the dendrogram. Panels $b$ and $c$ show two realizations of the model. The 'hierarchy' of a hierarchical random graph is not a hierarchy in the sense of a phylogenetic tree, but rather a way of encoding probabilities for links between each pair of nodes. Since the hierarchical random graph model is a collection of probabilities, each specific realization is different; on average the properties of the realized graphs reflect the properties of the model.

The hierarchical random graph model is very flexible. Generally, by selecting suitable inner probabilities, this model is able to capture most of the currently known network characteristics: degree distributions, degree-degree correlations, undirected motifs, and communities.

Clauset et al. use the hierarchical random graph model to gain insight into the structure of real networks. They begin with a real network and estimate what dendrogram (hierarchical random graph model) is most likely to have generated that particular network. The parameters of the hierarchical graph model contain condensed information about the actual network. The price of the flexibility of the hierarchical random graph model is a high number of variables that require fitting. In a model with $n$ nodes, this model requires that we fit *(n-1)* probabilities. The authors have found a robust way to solve this problem; they determine the correct model by performing Monte Carlo sampling of hierarchical random graphs with a probability proportional to the likelihood that the model results in the observed network.

The paper analyzes three real networks, the metabolic network of *Treponema pallidum*, a network of associations between terrorists, and a food web of grassland species. The sampled dendrograms from these networks yield new networks that are different in detail from the originals, but preserve important structural features of the original networks. In particular, the resampled networks match closely in terms of degree distribution, clustering coefficient, and distribution of path lengths to the original. Hence, the hierarchical structure of networks contains important information about other topological properties as well.

## Missing Links

The ability of the Clauset et al. algorithm to detect false positives and missing links makes it especially valuable in relation to biological systems. Despite the vast efforts and progress in high throughput experiments and mass spectrometry technologies, available cellular information is still sparse. Even in most studied organisms such as *Saccharomyces Cerevisiae*, the gene regulatory or the protein-protein interaction network is far from complete. In addition to gaps in data, every experimental method induces unavoidable biases into the data[35]. Low throughput experiments for protein-protein interaction measurements, for instance, tend to focus on well known proteins whereas high throughput experiments without quality control may produce false positives. The missing- or falsely existing links in the network naturally impact the correct identification of the functional modules.

Using the hierarchical random graph model and the associated hierarchy of link probabilities, the algorithm of Clauset et al. allows for identification of false positive and negative links. False positives are the links that exist despite the low link probability found by the method whereas false negatives are the links that do not exist in the network despite the high link probability. Nevertheless, the performance of this algorithm will naturally still depend on the signal to noise ratio in the data.

## THE FUTURE OF COMMUNITY DETECTION

With the explosion of available data, it has become clear that real networks can possess a variety of structural properties. Biological networks are distinct from social networks that are in turn quite distinct from information networks. The heterogeneity of different networks, however, poses a particular challenge for community detection. In many ways, community detection is like image segmentation. One of the primary reasons that segmentation tasks are difficult is because images can have (sometimes hierarchical) structure on many different scales. Consider the case of phase contrast microscopy. The task could be to isolate parts of the nucleus (nucleoli, mitochondria, etc), the nucleus itself, the whole cell, or even groups of cells with a particular phenotype. In order to efficiently detect an object in an image, it is important to know its scale and features.

Segmenting a cell on a background of many cells in a dish is markedly different than detecting a proboscis monkey in a tropical forest. In a similar manner, finding communities in social networks is quite a different task than detecting modules in cellular networks. Like image segmentation, community detection is more of an art form than a problem that we can blindly apply a brute-force algorithm to; skillful module identification requires knowledge of the subject matter and training.

## CHALLENGES

Complex network theory is a young field and its tools are constantly maturing. Here we discuss some avenues for future research. The algorithm by Clauset *et al*. takes an important step forward by being able to model a wide variety of network structures. It does so in an indirect manner, however. We need community detection algorithms that incorporate the known motifs directly in the statistical model. Most current models[16,17] make the assumption that networks are essentially some variation of a random graph, while we know that real networks are far from random on every level, e.g., certain motifs are much more likely than others.

We believe that a comprehensive module detection scheme should allow communities to overlap**Error! Bookmark not defined.**. Many proteins or genes are pleiotropic, and often associated with many functions. Hence a module detection algorithm that assigns proteins into several functional modules is biologically essential.

Another important factor to take into account is bipartite network information. A bipartite network is a network that contains two different types of nodes, and links run only between nodes of different types. Gene regulatory networks[12] (transcription factors and regulated genes) or protein-phosphorylation networks[21] (kinases and substrates) are bipartite. Many integrative biological networks are also bipartite, e.g., drug-target network[36] and the gene-disease network[37]. The usual way of analyzing bipartite networks is by projecting them onto one of the node sets, e.g., disease-disease network. However, it has been shown that this projection discards important network information[38]. An algorithm that incorporates bipartite information will allow for more refined community detection.

A final hurdle to cross has to do with validation. How do we know that the communities we found are the correct ones? How can we compare the results of two distinct clustering algorithms and declare that one is better than the other? Currently the state of the art is to design an artificial network with the structural properties that one wants to detect (e.g. group structure) and then show that the algorithm being tested is able to detect such structures. However, this process does not guarantee the performance on real networks. Ideally, we would like to compare the performance of two algorithms on a real data set. It is possible to artificially remove (or add) links from a real network and measure how well the algorithm under study is able to accurately determine robust community structure[39]. Algorithms can be compared based on performance during such tests.

Because of its flexibility and well-founded statistical nature, the algorithm suggested by Clauset et al. has the potential to encompass many of the challenges that face community detection.

ACKNOWLEDGEMENTS

Both authors contributed equally. We thank C. Hidalgo for his creative improvements of the figures and A.-L. Barabasi for valuable discussions. N.G. acknowledges support by National Institutes of Health (NIH) grant P50 HG004233 (M. Vidal, PI; subcontract to A.-L.B.). S.L. acknowledges support by the Danish Natural Science Research Council and James S. McDonnell Foundation 21st Century Initiative in Studying Complex Systems, the National Science Foundation within the DDDAS (CNS-0540348), ITR (DMR-0426737) and IIS-0513650 programs, as well as by the U.S. Office of Naval Research Award N00014-07-C and the NAP Project sponsored by the National Office for Research and Technology (KCKHA005).

## REFERENCES

[1]Clauset A, Moore C, Newman MEJ. 2008. Hierarchical structure and the prediction of missing links in networks. Nature 453: 98.

[2]Barabási A-L, Albert R. 1999. Emergence of Scaling in Random Networks. Science 286: 509.

[3]Rual J-F, Venkatesan K, Hao T, Hirozane-Kishikawa T, Dricot A, et al. 2005. Towards a proteome-scale map of the human protein–protein interaction network. Nature 437: 1173.

[4]Stelzl U, Worm U, Lalowski M, Haenig C, Brembeck F, et al. 2005. A Human Protein-Protein Interaction Network: A Resource for Annotating the Proteome. Cell 122: 957-968.

[5]Newman MEJ, Barabasi A-L, Watts DJ. 2006. The Structure and Dynamics of Networks: (Princeton Studies in Complexity). Princeton University Press.

[6]Caldarelli G. 2007. Scale-Free Networks: Complex Webs in Nature and Technology (Oxford University Press, USA).

[7]Boccaletti S, Latora V, Moreno Y, Chavez M, Hwang D-U. 2006. Complex networks: Structure and dynamics. Physics Reports 424:175.

[8]Dorogovtsev SN, Goltsev AV, Mendes JFF. 2007. Critical phenomena in complex networks. http://arxiv.org/abs/0705.0010. Accepted in Rev. Mod. Phys.

[9]Pastor-Satorras R, Vespignani A. 2004. Evolution and Structure of the Internet: A Statistical Physics Approach (Cambridge University Press).

[10]Maslov S, Sneppen K. 2002. Specificity and Stability in Topology of Protein Networks. Science 296: 910.

[11]Gonzalez MC, Herrmann HJ, Kertesz J, Vicsek T. 2007. Community structure and ethnic preferences in school friendship networks. Physica A 379: 307-316.

[12]Milo R, Shen-Orr S, Itzkovitz S, Kashtan N, Chklovskii D, Alon U. 2002. Network Motifs: Simple Building Blocks of Complex Networks. Science 298: 824-827.

[13]Watts DJ, Strogatz S. 1998. Collective dynamics of small-world networks. Nature 393: 440–442.

[14]Bianconi G, Marsili M. 2005. Loops of any size and Hamilton cycles in random scale-free networks. J. Stat. Mech.-Theor. Exp. P06005.

[15]Bianconi G, Gulbahce N, Motter AE. 2008. Local Structure of Directed Networks. Phys. Rev. Lett. 100: 118701.

[16]Fortunato S, Castellano C. 2009. Community Structure in Graphs. In Encyclopedia of Complexity and Systems Science (Springer).

[17]Danon L, Díaz-Guilera A, Duch J, Arenas A. 2005. Comparing community structure identification. J. Stat. Mech. -Theor. Exp. P09008.

[18]Newman MEJ. 2006. Modularity and community structure in networks. PNAS 103: 8577.

[19]Arenas A, Diaz-Guilera A, Perez-Vicente CJ. 2006. Synchronization Reveals Topological Scales in Complex Networks. Phys. Rev. Lett. 96: 114102.

[20]Ravasz E, Somera AL, Mongru DA, Oltvai ZN, Barabasi A-L. 2002. Hierarchical Organization of Modularity in Metabolic Networks. Science 297: 1551-1555.

[21]Linding R, Jense LJ, Ostheimer GJ, van Vugt MATM, Jorgensen C, et al. 2007. Systematic discovery of in vivo phosphorylation networks. Cell 129: 1415-1426.

[22]Giaever G, Chu AM, Ni L, Connelly C, Riles L. 2002. Functional profiling of the *Saccharomyces cerevisiae* genome. Nature 418: 387-391.

[23]Tong AHY, Evangelista M, Parsons AB, Xu H, Bader GD. 2001. Systematic Genetic Analysis with Ordered Arrays of Yeast Deletion Mutants. Science 294: 2364-2368.

[24]Tong AHY, Lesage G, Bader GD, Ding H, Xu J et al. 2004. Global Mapping of the Yeast Genetic Interaction network. Science 303: 808-813.

[25]Motter AE, Gulbahce N, Almaas E, Barabasi A-L. 2008. Predicting Synthetic rescues in Metabolic Networks. Mol. Sys. Bio. 4: 168.

[26]Palla G, Derenyi I, Farkas I, Vicsek T. 2005. Uncovering the overlapping community structure of complex networks in nature and society. Nature 435: 814.

[27]Rosvall M, Bergstrom CT. 2007. An information-theoretic framework for resolving community structure in complex networks. PNAS 104: 7327.

[28]Frey BJ, Dueck D. 2007. Clustering by Passing Messages between Data Points. Science 315: 972.

[29]Hofman JM, Wiggins CH. 2007. A Bayesian Approach to Network Modularity. http://arxiv.org/abs/0709.3512.

[30]Newman MEJ, Leicht EA. 2007. Mixture models and exploratory analysis in networks. PNAS 104: 9564.

[31]Newman MEJ, Girvan M. 2004. Finding and evaluating community structure in networks. Phys. Rev. E 69: 026113.

[32]Johnson SC. 1967. Hierarchical Clustering Schemes. Psychometrika, 2:241-254.

[33]Eisen MB, Spellman PT, Brown PO, Botstein D. 1998. Cluster analysis and display of genome-wide expression patterns. PNAS 95:14863-14868.

[34]Barabasi A-L, Ravasz E, Vicsek T. 2001. Deterministic scale-free networks. Physica A 299: 559.

[35]Vidal M. 2001. A biological atlas of functional maps. Cell 104: 333.

[36]Yildirim MA, Goh K-Il, Cusick ME, Barabasi A-L, Vidal M. 2007. Drug-target network. Nature Biotech. 25: 1119.

[37]Goh K-Il, Cusick ME, Valle D, Childs B, Vidal M, Barabasi A-L. 2007. The human disease network. PNAS 104: 8685.

[38]Lehmann S, Schwartz M, Hansen LK. 2008. Bi-clique Communities. http://arxiv.org/abs/0710.4867 (2007). Accepted in Phys. Rev. E.

[39]Karrer B, Levina E, Newman MEJ. 2008. Robustness of community structure in networks. Phys. Rev. E 77: 046119.